\documentclass[lettersize,journal]{IEEEtran}
\usepackage{amsmath,amsfonts}
\usepackage{algorithmic}
\usepackage{algorithm}
\usepackage{array}
\usepackage[caption=false,font=normalsize,labelfont=sf,textfont=sf]{subfig}
\usepackage{textcomp}
\usepackage{stfloats}
\usepackage{url}
\usepackage{verbatim}
\usepackage{graphicx}
\usepackage{cite}
\usepackage [english]{babel}
\usepackage [autostyle, english = american]{csquotes}
\MakeOuterQuote{"}

\begin{document}

\title{The Simons Observatory: Differentiating Thermal and Optical Effects in Superconducting Transition-Edge Sensing Bolometers}

\author{Rita F. Sonka, Shannon M. Duff, Daniel Dutcher, Suzanne T. Staggs

\thanks{This work was supported in part by a grant from the Simons Foundation (Award 457687, B.K.). Simons Observatory is funded by the Simons Foundation, the Heising-Simons Foundation, and other research institutions within the collaboration. \textit{(Corresponding author: Rita F. Sonka)}}
\thanks{R. F. Sonka, D. Dutcher, and S. T. Staggs are with Princeton University Physics, Jadwin Hall, Princeton, NJ 08544-2013 USA (e-mail: rfsonka@gmail.com; ddutcher@princeton.edu; staggs@princeton.edu).}
\thanks{S. M. Duff is with National Institute of Standards and Technology, 325 Broadway, Boulder CO 80305 USA (e-mail: shannon.duff@nist.gov).}
}

\maketitle

\begin{abstract}

The Simons Observatory aims to field 70,000 Transition-Edge Sensor (TES) bolometers to  measure the Cosmic Microwave Background. With so many detectors, rapid but accurate validation of their properties prior to their integration into telescopes is of particular importance. This paper describes an exploration of a new method to improve the simultaneous characterization of TES thermal parameters and bolometer optical efficiencies without significantly increasing the data collection time.  The paper uses a special-purpose data set comprising current-voltage (IV) curves collected from thousands of TES bolometers with a variety of different average bath temperatures and different cold load temperatures. A subset of the bolometers were masked so they received no optical power.  The new method fits data from the bath temperature ramp and cold load temperature ramps together as one set instead of fitting each independently. This enables thermal parameter assessment of the unmasked detectors without performing additional cooldowns of the cryostat, halving the time necessary to obtain thermal characterization of all detectors.  

\end{abstract}

\begin{IEEEkeywords}
Microwave detectors, Satellites and large arrays, Superconducting Detectors, Superconducting device testing, Temperature measurement, Thermal properties, Transition-edge sensors (TES) devices 
\end{IEEEkeywords}

\section{Motivation}

\IEEEPARstart{T}{he} Simons Observatory (SO) aims to observe, map and analyze the cosmic microwave background (CMB) in order to characterize the primordial perturbations (B-modes), measure the number of relativistic species and the mass of neutrinos, test for deviations from a cosmological constant, improve our understanding of galaxy evolution, and constrain the duration of reionization \cite{Ade_2019}. 

\begin{figure}[!t]
\centering
\includegraphics[width=3in]{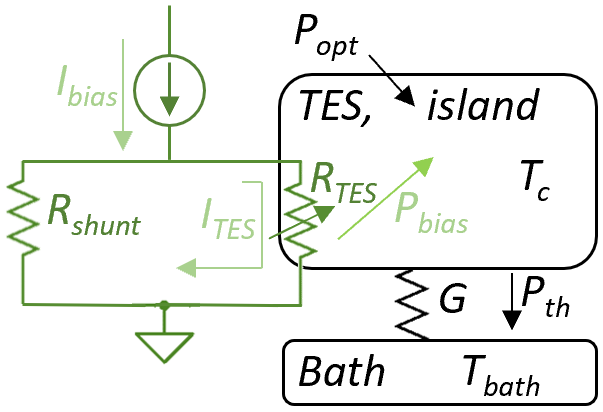} 
\caption{Diagram of the ideal TES coupled thermal (black) and electrical (green) circuit.}
\label{fig:detector_background}
\end{figure}

Due to the signal's tiny amplitude \cite{abazajian2019cmb}, SO must fabricate, characterize and deploy tens of thousands of optically-coupled, densely-packed Transition-Edge Sensors (TESes) in their telescopes to meet the science goals. SO will deploy $\sim$70,000 detectors packaged within 49 detector-readout focal plane modules distributed among four telescopes \cite{mccarrick_2021}.
This paper describes a new method for simultaneously  characterizing the detector's thermal parameters and optical efficiencies without requiring more than one time-consuming cooldown of the devices. The method could be used by other instruments that use TES bolometers.

\section{Background: TES Operation and Parameters}

Fig.\,\ref{fig:detector_background} diagrams the coupled thermal and electrical circuit of a TES bolometer. In operation, $I_{bias}$, the current along the bias line, is held constant. Since the shunt resistance, $R_{shunt}$, is less than the (variable) TES resistance, $R_{TES}$, within the TES superconducting transition, this effectively voltage-biases the TES. This combined with the thermal circuit creates negative electrothermal feedback that keeps the TES within the superconducting transition even as $P_{opt}$, the optical power deposited on the bolometer, varies (until it is high enough to drive the TES normal)\cite{irwin_2005}. The current through the TES, $I_{TES}$, is read out through a microwave multiplexed readout system (not pictured) \cite{mates2011microwave} \cite{mccarrick2021simons}. The electrical power dissipated in the TES, $P_{bias}$, can then be calculated from $I_{bias}$, $I_{TES}$, and $R_{shunt}$. From there, $P_{opt}$, the desired measurable quantity, can be calculated \cite{irwin_2005} \cite{sudiwala2002thermal} from:

\begin{figure}[!t]
\centering
\includegraphics[width=3.5in]{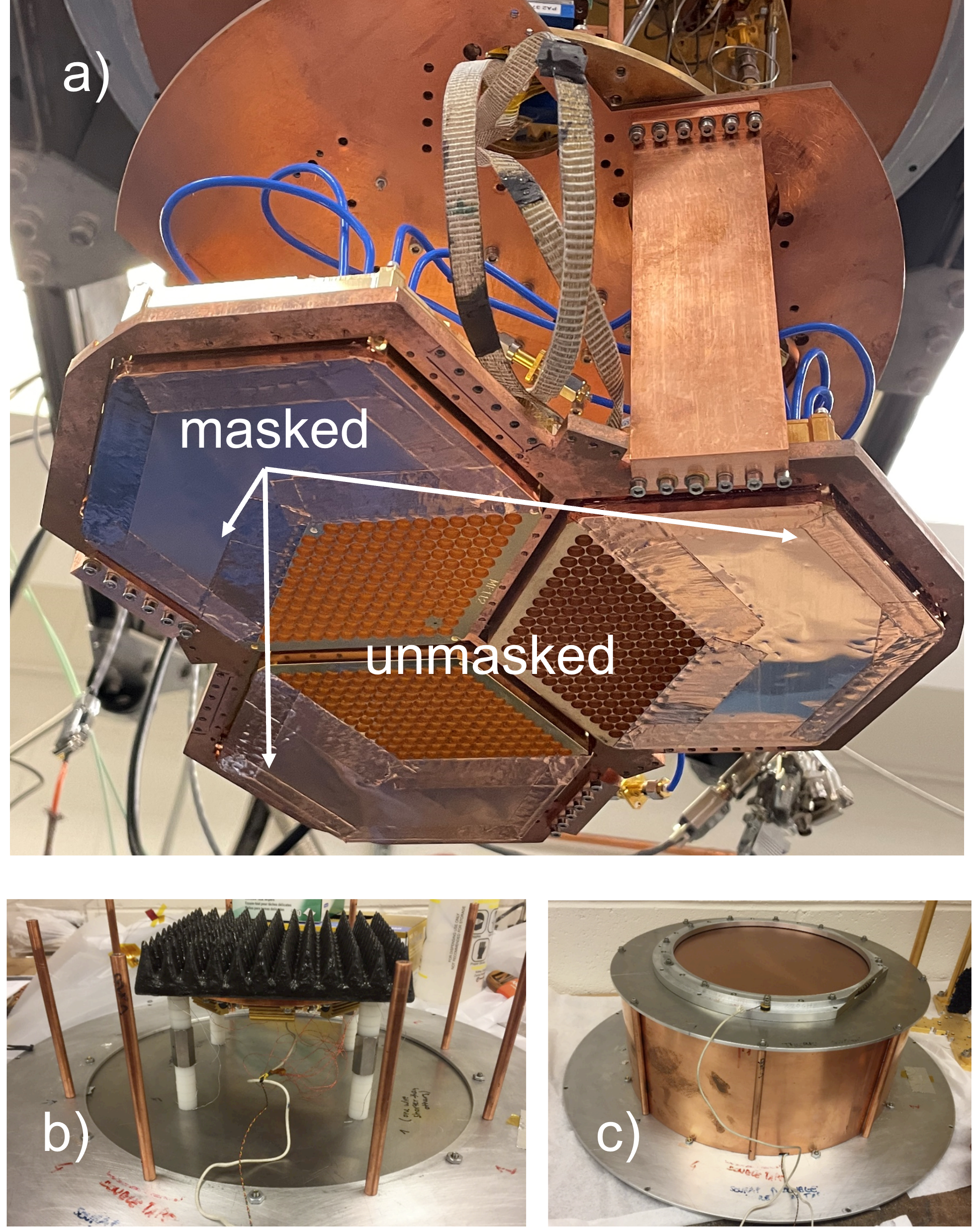}
\caption{Experimental setup illustrations. The top photograph (a) shows a sky-side view of three focal plane modules installed in the cryostat for cooldown. Their optical feedhorns point downwards, towards the cold load \cite{wang_2022}. Two-thirds of the feedhorns are masked with copper shim stock, preventing them from receiving light. The bottom left photograph (b) shows the impedance-matched Eccosorb CR114 cold load outside of its optical filter and thermal containment casing \cite{li_2020probing}. The bottom right photograph (c) shows the cold load inside thermal containment casing with its bandpass optical filter on top. The cold load is  mounted beneath the modules, and is weakly linked to the 4K stage of the cryostat so its temperature can be ramped from 9.5 K to 18 K.  The resulting optical power for each detector is estimated from the detector beams and the cold load geometry (see \cite{Choi_2018} for similar work).}
\label{fig:experimental_setup}
\end{figure}

\begin{equation}
\begin{split}
    P_{th} = \frac{G}{n T_{c}^{n-1}}  (T_{c}^{n} - T_{bath}^{n}) 
    = P_{b70} + \eta_{opt} P_{opt} 
\end{split}
\label{eq:p_th}
\end{equation}

where $P_{th}$ is the total thermal power vented into the bath; $P_{b70}$ is $P_{bias}$ at the point in the transition where $R_{TES} = 0.70 R_n$, with $R_n$ the normal resistance of the TES;  $G$ is the differential thermal conductance connecting the TES island and bath; 
$T_c$ is the TES critical temperature, the temperature at which the TES transitions between normal resistance and superconductivity; 
$n$ is the power law index (1 + the thermal conductance exponent); and 
$\eta_{opt}$ is the optical efficiency, the ratio of how much power the cold load delivers to the bolometer's coupled orthomode transducer (OMT) waveguide to how much power it absorbs.

The three TES thermal parameters ($G$, $T_c$, $n$) and the TES optical efficiency ($\eta_{opt}$) must be obtained in the laboratory before the detectors are fielded to confirm the detectors will meet noise specifications and to solve Eq.\,\ref{eq:p_th} for $P_{opt}$.

\section{Datasets for Obtaining TES Thermal Parameters and Optical Efficiency}

\begin{figure}[!t]
\centering
\includegraphics[width=3.5in]{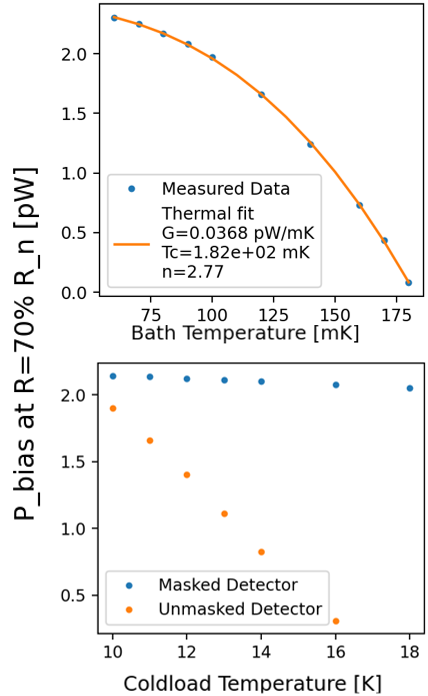} 
\caption{Independent fitting techniques examples. The top graph shows the bath ramp data and thermal fit results for one masked 90 GHz detector. The bottom graph shows that masked detector's cold load ramp data (at the top) and the cold load ramp data for an unmasked 90 GHz detector positioned at a similar distance from the cold load center. Notice that the masked detector's $P_{b70}$s do decrease slightly with increasing cold load temperature; this is due to indirect thermal heating of the bath in that area from the cold load power. } 
\label{fig:original_fitting_technique}
\end{figure}

As illustrated in Fig.\,\ref{fig:experimental_setup}, the detectors were installed in a dilution refrigerator (DR) above a cold load made of Eccosorb CR114 \cite{wang_2022}. Two-thirds of the detectors were masked, preventing light from reaching them, while the rest were left unmasked to receive power from the cold load.
After the detectors cooled below 100 mK, the following two data sets were taken (see Fig.\,\ref{fig:original_fitting_technique} for example data from two detectors) over the course of about 8 hours each:

\begin{enumerate} 
\item{"Bath ramp" - the bath temperature $T_b$ was repeatedly increased and allowed to reach equilibrium, while the cold load was kept at 9.5 K. At each bath temperature, current vs. voltage (IV) curves (see \cite{Gildemeister_2000}) for all the detectors were taken.}
\item{"Cold load ramp" - the temperature of the cold load was repeatedly increased and allowed to reach equilibrium, while the bath temperature was feedback controlled to 80 mK, referenced off a thermometer on the copper module mount. At each cold load temperature, IV curves for all the detectors were taken. Also at each cold load temperature, the $P_{opt}$ on every position on the module was estimated from the thermal design and geometry of the cryostat and cold load \cite{Choi_2018}.}
\end{enumerate}

\section{Fitting Techniques}
\subsection{The Independent Fits Technique}

A $P_{b70}$  was extracted from each IV curve. Then Eq.\,\ref{eq:p_th} was fit to each masked ($P_{opt}=0$) detector's bath ramp data (only) to obtain its thermal parameters, as illustrated in the top plot of Fig.\,\ref{fig:original_fitting_technique}.  Typically,  $\eta_{opt}$ for each unmasked detector would then be extracted from the cold load ramp data, using the masked detectors to correct for any wafer heating associated with changing the temperature of the cold load (see ex. \cite{Choi_2018}). 

\subsection{The Together-fits Technique}

\begin{figure}[!t]
\centering
\includegraphics[width=3.5in]{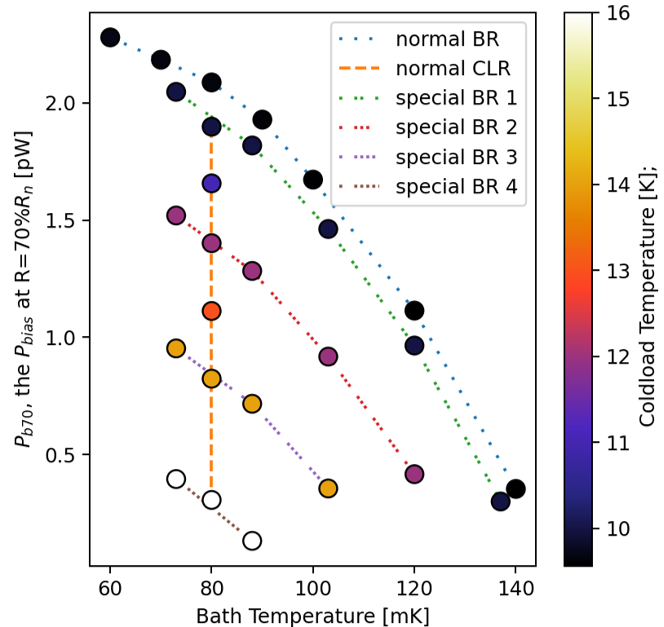} 
\caption{Example of applying the together-fits technique to an unmasked 90 GHz detector in the special exploration data set. The standard data set would only include the normal bath ramp and normal coldload ramp. BR= bath ramp, CLR = Cold Load Ramp.} 
\label{fig:special_dataset}
\end{figure}

\begin{figure*}[!t]
\centering
\subfloat[Original bath ramp, All detectors fit as if masked]{\includegraphics[width=3.5in]{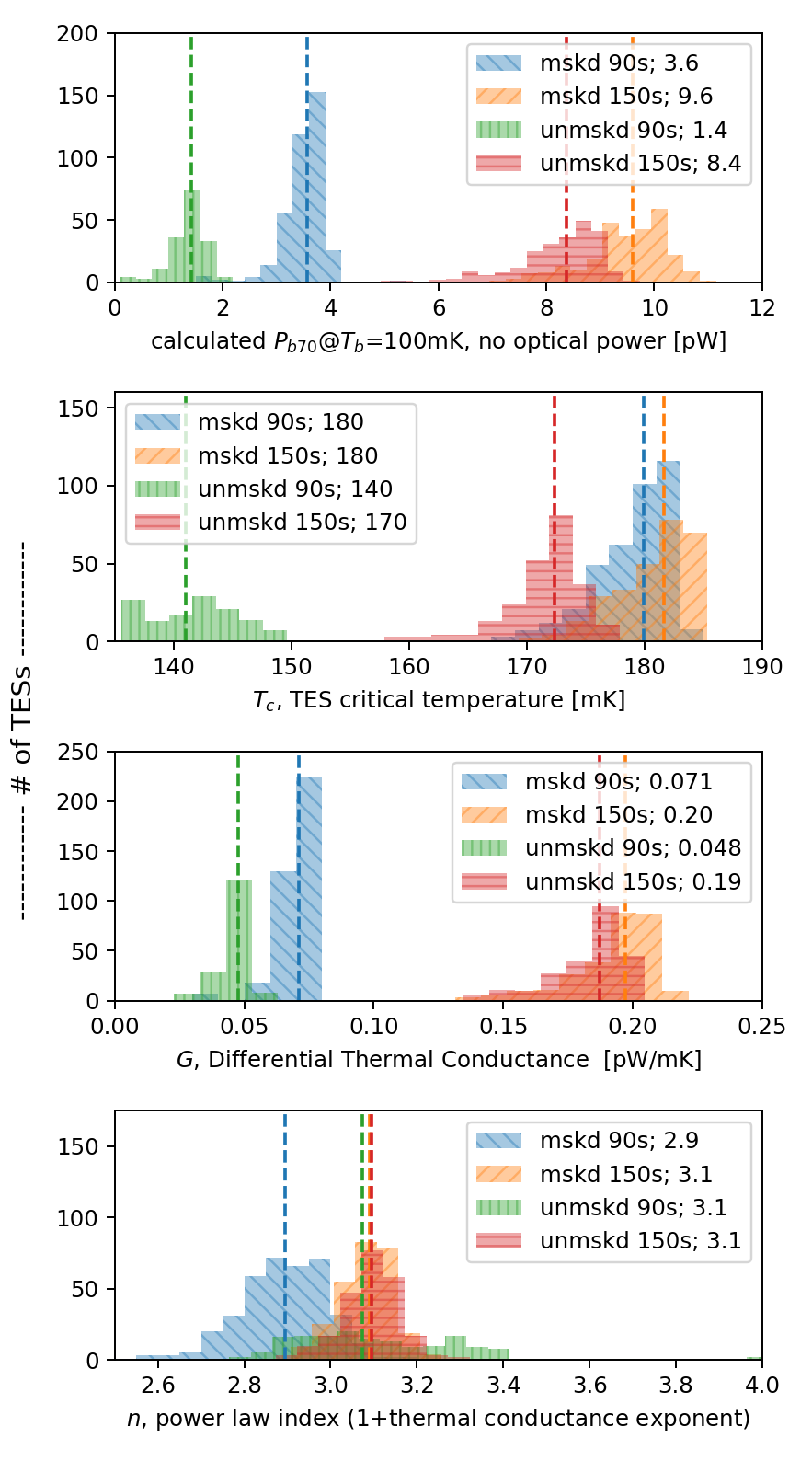}
\label{fig:bathramp_histograms}}
\hfil
\subfloat[Together-fits for those detectors]{\includegraphics[width=3.5in]{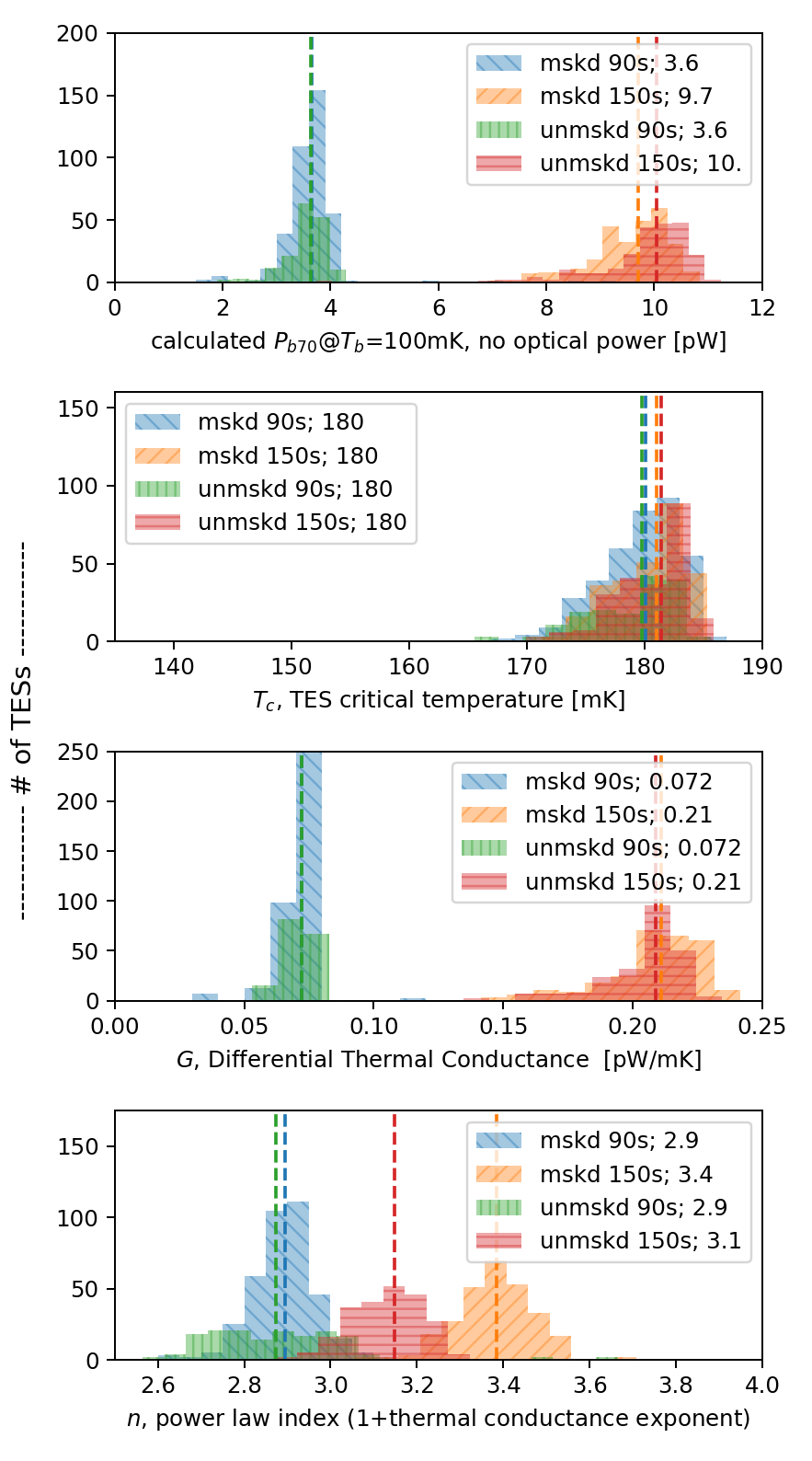}
\label{fig:together_fits_histograms}}
\caption{Histograms of the thermal parameter fits for detectors in one module (designated Mv12) using (a) the  standard method,  and (b) together-fits methods on the special dataset. All detectors fit by both methods are included. In the legends, "unmskd"= unmasked, "mskd"=masked, "90s"=90 GHz frequency detectors,  "150s"=150 GHz detectors, and the number at the end is the median, in the units given in the x-axis label. The vertical dashed lines display the medians on the plots.   Note the bimodal distributions of the thermal parameters visible in (a), that come from fitting the unmasked detectors. (b) illustrates how the together-fits technique obtains proper thermal fits of these detectors.} 
\label{fig:histograms}
\end{figure*}

The idea of the together-fits technique (see Fig.\,\ref{fig:special_dataset}) is to fit the bath ramp and cold load ramp for a given detector together, with a new parameter, a dark heating coefficient,  $\xi_{therm}$: 

\begin{equation}
    P_{b70}  = \frac{G}{n T_{c}^{n-1}} (T_{c}^{n} - T_{bath}^{n}) - \eta_{opt} P_{opt} - \xi_{therm} P_{opt}.
\label{eq:p_tf}
\end{equation}

Although masked detectors' OMTs do not absorb optical power directly (so their apparent $\eta_{opt} = 0$),  their $P_{b70}$ values do change when the cold load temperature changes due to parasitic heating from the cold load altering their individual bath temperatures \cite{Choi_2018}. The bath ramp and cold load ramp data for dark detectors are simultaneously fit for $\xi_{therm}$, $G$, $n$, and $T_c$. 

Then, for the optical detectors, the bath ramp and cold load ramp data are fitted for $G$, $n$, $T_c$, and $\eta_{opt}$  with $\xi_{therm}$ fixed as the average of the $\xi_{therm}$ values for the masked detectors at similar distance from the cold load center.  

\subsection{Advantages of the Together-fits Technique}

This technique enables calculation of thermal parameters of unmasked detectors, and provides a more accurate estimate of the masked detectors' thermal parameters.   Estimates of both $\eta_{opt}$ and $\xi_{therm}$ for all the detectors in a module can be used to disambiguate signal fluctuations due to atmospheric fluctuations in the field from those caused by bath temperature fluctuations.  
Additionally, it is useful for predicting on-sky simultaneous biasability of detectors associated with each of the 12 module bias lines \cite{mccarrick_2021, sonka_2017characterization} because it allows incorporation of the impact of $\eta_{opt}$ variations among the detectors.  

\section{Exploratory Data set and Example Results}
To explore the power and validity of this technique, a special data set was taken in addition to the normal bath ramp and cold load ramp. It consisted of four smaller bath ramps, each at a different elevated cold load temperature; example results for one detector are shown in Fig.\,\ref{fig:special_dataset}. The overlap of the cold load ramp-derived $P_{b70}$s with the different bath ramp $P_{b70}$s confirms that the method of approaching a given bath temperature/cold load temperature phase space point does not significantly affect the measurement. 

Fig.\,\ref{fig:histograms} tests and showcases the together-fits technique. 
The four histograms on the left side show the results of fitting the thermal parameters for all the detectors in one module with the standard SO method (normally only applied to the masked detectors), which relies on a single bath ramp data set at fixed cold load temperature. 
The histograms include both masked and unmasked detectors.  As expected, the $P_{b70}$ values are smaller for the unmasked detectors.  

The four histograms on the right side derive from the together-fits method on the same detectors. Note that now the masked and unmasked detectors show much better agreement for $P_{b70}$ and $G$. The convergence of $T_c$ most strongly signals the accuracy, as $T_c$ is the thermal parameter fabricated most consistently among the detectors (barring a potential small radial variation). This is because it is intended to be the same for all of them, and is finely tuned simultaneously for the entire wafer by a heating process \cite{li2016AlMn} (and thus cannot suffer midway through the module fabrication from degradation in mechanical precision or operator attentiveness, as the other parameters can.). The difference in average 
 fitted $T_c$ between masked and unmasked detectors is 21 in the original method; 1.0 (95\% decrease) when the together-fits method is applied to just the normal dataset; and 0.66 (97\% decrease) when the together-fits method is applied to the special dataset.

\section{Going Forward}

The next step for capitalizing on this together-fits method is to define a new way of collecting test data for SO modules that improves the fitting power without taking significantly longer than the standard method ($\sim 8$ hrs). We will analyze the special dataset (which took $\sim 36$ hrs to acquire) with more flexible models, such as allowing $\chi_{therm}$ to vary with bath temperature. With those results in mind, we will further analyze the special dataset to select the optimal set of (bath-temperature, cold load-temperature) points consistent with the time constraint. Note that applying the together-fits technique to just the first bath ramp and cold load ramp (not including the extra data in the special dataset) does still extract much more accurate thermal parameter values for the unmasked detectors than the independent-fits method. 
Further work will also explore use of  the 36 `dark' bolometers  fabricated on the SO detector wafers without any optical coupling, and other ways of estimating the dark heating term, and its usefulness in isolating bath temperature fluctuations from other signal variations in the field.  

\bibliography{Together_fits_manuscript}
\bibliographystyle{IEEEtran}

\vfill

\end{document}